
\def\rie{ R_{\mu\nu}}
\def\ez{e^{2\zeta}}

\def\section#1{\bigskip\noindent{\bf#1}\smallskip}

\magnification=1200

\font\titolino=cmbx10
\font\tsnorm=cmr10
\font\tscors=cmti10

\font\tscorsp=cmti9
\magnification=1200

\def\PRD{{\tscors Phys. Rev. D }}

\def\NPB{{\tscors Nucl. Phys. B }}

\def\PLB{{\tscors Phys. Lett. B }}

\def\PRP{{\tscors Phys. Rep. }}

\def\MPLA{{\tscors Mod. Phys. Lett. A  }}

\def\ANP{{\tscors Ann. Physics} (N.Y.) }
\def\CMP{{\tscors Commun. Math. Phys. }}

\def\note{\advance\notenumber by 1 \footnote{$^{\the\notenumber}$}}
\def\ref#1{\medskip\everypar={\hangindent 2\parindent}#1}
\def\beginref{\begingroup
\bigskip
\leftline{\titolino References.}
\nobreak\noindent}
\def\endref{\par\endgroup}
\def\beginsection #1. #2.
{\bigskip
\leftline{\titolino #1. #2.}
\nobreak\noindent}
\def\beginack
{\bigskip
\leftline{\titolino Acknowledgments.}
\nobreak\noindent}

\nopagenumbers
\null
\vskip 5truemm
\rightline {INFNCA-TH9523}

\rightline { }
\rightline { }
\rightline{November 1995}
\vskip 15truemm
\centerline{\titolino DILATONIC BLACK HOLES IN A S-DUALITY MODEL}
\vskip 15truemm
\centerline{\tsnorm S. Monni and  M. Cadoni}
\bigskip
\centerline{\tscorsp Dipartimento di Scienze Fisiche,}
\smallskip
\centerline{\tscorsp Universit\`a  di Cagliari,}
\smallskip
\centerline{\tscorsp Via Ospedale 72, I-09100 Cagliari, Italy.}
\smallskip
\centerline {\tscorsp and}
\smallskip
\centerline{\tscorsp INFN, Sezione di Cagliari.}
\bigskip
\vskip 15truemm
\centerline{\tsnorm ABSTRACT}
\begingroup\tsnorm\noindent
\baselineskip=2\normalbaselineskip
We find exact charged black  hole solutions of a string effective action
that is invariant under S-duality transformations. These black hole
solutions have the same causal structure as the Reissner-Nordstrom (RN)
solutions. They reduce to the RN solutions for self-dual configurations of the
dilaton and to the Garfinkle-Horowitz-Strominger (GHS) solution in the weak
(or strong) coupling regime. Using the purely magnetic solutions of the
S-duality model we  also
generate dyonic black hole solutions of the GHS model, which have the
causal structure of the RN solutions.


\vfill
\leftline{\tsnorm PACS: 04.70.Bw, 11.25.-W\hfill}
\smallskip
\hrule
\noindent
\leftline{E-Mail: CADONI@CA.INFN.IT\hfill}
\bigskip
\endgroup
\vfill
\eject
\footline{\hfill\folio\hfill}
\pageno=1
\beginsection 1. Introduction.
\smallskip

One feature of four-dimensional string effective theory is the
existence of charged black hole solutions that are drastically different
from the Reissner-Nordstrom (RN) solutions of general relativity [1]. String
theory predicts the existence of scalar fields, such as the dilaton and
the moduli, whose couplings to the Maxwell field $F_{\mu\nu}$ enable one
to circumvent the uniqueness  and no hair-theorem stating that the
only static black hole solution to the Einstein-Maxwell equations is the
RN solution [2]. The coupling of the dilaton and the moduli
to the Maxwell field is parameterized by a coupling function $f$, which
therefore determines the strength of the gauge couplings. At the tree
level in the string perturbation theory (spherical worldsheet topology)
 the function $f$ depends,
exponentially,  only on the dilaton [3]. The charged black hole solutions of
the tree-level string action have been found by Gibbons and Maeda [4] and
later
rediscovered by Garfinkle, Horowitz and Strominger (GHS) [1].
These black hole  solutions have features that make
them drastically different from the RN solutions. They have an event
horizon but no inner (Cauchy) horizon and in the extremal limit the area
of the event horizon goes to zero giving zero entropy. In this limit, for a
magnetically charged black hole, the event horizon moves off to infinity
in all directions differently from the RN case.

This new framework for the description of charged black holes poses new
questions and problems on the subject, which have been extensively
debated in the literature [5]. The low-energy string effective action used
by the GHS model is just a first approximation that holds at the tree
level in the string perturbation theory and at the leading order in the
inverse string tension $\alpha'$. Perturbative corrections to the coupling
function $f$ and/or nonperturbative effects  may, in principle,  change
the description of the charged
black holes in string theory. The basic question involved here is as
follows: Is the above description of charged black holes a general feature
of string theory or is it just an artifact of the particular approximation?
Progress in this direction has been made in ref. [6,7], but a general answer
to this question is still lacking.
A second and related problem is the relationship between the string and
the RN description of charged black holes. Because string effective
theory reduces to general relativity in the region of weak string
couplings,
where the dilaton $\Phi$ is approximately constant,  one would expect
there the RN
solutions to be a
good approximation to string charged black holes. However, the purely
magnetic (or purely electric) GHS black
hole does not reduce in any approximation to the RN solution. This fact
led the authors of ref. [1] to the conclusion that the RN solution is not
even an approximate solution of string theory. This behavior can be
traced back to the fact that in the GHS model the dilaton has an
exponential coupling to $F^2$,  so every solution with $F^2\neq 0$ must
have a nonconstant dilaton. Again, this is true only for
$f=\exp(-2\Phi)$ and for purely electric (or purely magnetic)
configurations,  with a different choice for the coupling function $f$ or
for dyonic configurations
the situation could change drastically. Dyonic solutions  of the GHS model,
 which have the causal
structure of the RN black hole,
have been already found by Gibbons and Maeda [4].

In this paper we will tackle the previous problems by studying a model
with a coupling function $f=\cosh(2g\Sigma)$. If one interprets the field
$\Sigma$ as the dilaton, the model can be viewed as a way to implement the
S-duality symmetry, which has been conjectured to hold in string theory [8],
in the context of a low-energy string effective action.
 On the other
hand, if one considers $\Sigma$ as a modulus, the same model can be thought
as an approximation to the effective action resulting from toroidal or
orbifolds
compactifications. In this case the effective action  is known to be invariant
under the T-duality
symmetry $\Sigma \to-\Sigma$ [9].
We will find exact charged black hole solutions for the models with
$g^2=1,3$. These black hole solutions have the same Carter-Penrose diagram
as the RN solutions. In particular, the solutions with $g^2=1$ differ from
RN case just in the areas of the spheres with $r=const, t=const$. Moreover,
the following  feature of these black holes will emerge:
they reduce to the RN solutions for  self-dual configurations of the field
$\Sigma$ and to the GHS solution in the weak (or strong) coupling
regime.
Using the purely magnetic solutions of the S-duality model we will also be
able to
generate dyonic black hole solutions of the GHS model, which again have the
causal structure of the RN solutions.

The outline of the paper is the following. In sect. 2 we describe the
model we  investigate. In sect. 3 we solve the field equations  of
the theory by reducing them to an  equivalent, Toda molecule, dynamical
system and we
analyze in detail the corresponding charged black hole solutions. In
sect. 4 we use our model to generate dyonic black hole  solutions of
the GHS model.
Finally, in sect. 5 we present our conclusions.

\beginsection 2. The model.
\smallskip

We shall consider a model  described by the following action:
$$A=\int d^4x\sqrt{-g}\bigl(R-2(\nabla\Sigma)^2-\cosh(2 g\Sigma)
F^2\bigr),\eqno(2.1)$$
where $R$ is the scalar curvature, $\Sigma$ a
scalar field,
$F$ the abelian gauge  field strength and $g$ is a coupling constant.
If one considers $\Sigma$ as the dilaton, this action differs from the usual
low-energy 4D string action in
the choice of
coupling function $f(\Sigma)$ in the gauge field kinetic term $f(\Sigma)F^2$.
At the tree level in the string perturbation theory  $f=\exp(-2\Sigma)$ [3].
It has been conjectured that string theory is invariant
under a discrete $SL(2,Z)$ symmetry  associated with the field
${S=\exp(-2\Sigma) + i\Theta}$,  $\Theta$ being the axion field [8].
In particular this
invariance includes a symmetry $\Sigma \to -\Sigma$,
 which
exchanges weak and strong string couplings, $g^2_s \to 1/ g^2_s$.
In this context the model (2.1), which uses $f=\cosh(2g\Sigma),$
 may be viewed  as an S-duality invariant
modification of the tree-level dilaton coupling function
$f=\exp(-2\Sigma)$.

One can also interpret $\Sigma$ as a
modulus field associated with an overall radius of compactification. In
this case the action (2.1)  represents a T-duality invariant model of the
type considered in ref. [7].
It turns
out that for
toroidal compactifications and a large class of orbifolds, the coupling
function,
at the one-loop level in the string perturbation theory, can be split into the
sum of
the tree-level
dilaton-dependent
 gauge kinetic function  and a modulus-dependent term [9]:

 $$f= e^{-2\Phi}+a\ln\bigl(|\eta(T)|^4(T+T^*)\bigr) +b,\eqno(2.2) $$
 where $T=\exp(2\Sigma/\sqrt{3})$, $\eta(T)$ is the Dedekind function,
  $\Phi$ is the dilaton
  and $a$,$b$  are some constants.
 In particular, the genus-one
threshold correction term is invariant under the duality symmetry ($\Sigma
\to -\Sigma$). If one decides to study the strong
coupling region ($ \Phi \to \infty$) and uses
 $\cosh(2g\Sigma) $ as an
approximation to the  function $f$ given by eq. (2.2),
the function (2.2) reduces to the one in (2.1). In this case it is necessary
 to
introduce in the action (2.1) a
kinetic  term for the dilaton. This term does not  modify
the solutions since the dilaton is uncoupled and, due to the no-hair theorem,
it is constant
(obviously, consistency requires this constant to be chosen very large).
The solutions we are going to find will be examples of modulus
solutions in a curved spacetime and they can be considered as an extension
of some solutions previously studied  in the flat space case [7].

\beginsection 3. Black hole solutions.
\smallskip

The field equations stemming from the action (2.1) are:
$$\eqalign{\rie &=2\nabla_\mu
\Sigma\nabla_\nu\Sigma+2\cosh(2g \Sigma)\left(F_{\mu\rho}F_\nu^\rho-
{1\over 4}F^2g_{\mu\nu}\right),\cr\nabla^2\Sigma &={g\over
2}\sinh(2g\Sigma) F^2,\cr
\nabla_\mu&\bigl(cosh(2g\Sigma)F^{\mu\nu}\bigr)=0.\cr}\eqno (3.1)$$
Spherically symmetric solutions  of these equations can be found
using an ansatz that reduces
the system to a Toda-lattice form [4]:
$$\eqalign{d s^2&=e^{2\nu}(-d
t^2+e^{4\rho}d\xi^2)+e^{2\rho}d\Omega^2,\cr
F&=Q\sin\theta d\theta \wedge d\varphi,\cr}\eqno(3.2)$$
where $\nu$ and $\rho$ are functions of $\xi$ and $Q$ is the magnetic
charge.
We consider here only magnetic monopole configurations for the
electromagnetic (EM) field. Our results can be easily generalized to a purely
electric configuration using the invariance of the field equations
(3.1) under the EM duality transformation [10]:
$$f\to f^{-1}, \quad F\to fF^*,\quad \Sigma \to \Sigma,$$
with $f=\cosh(2g \Sigma)$.
The magnetic solutions of the theory with coupling function $f$ are related
 to the
electric solutions of the theory with coupling function $f^{-1}$.

Defining $\zeta=\nu+\rho$ and using the ansatz (3.2), the field
equations (3.1)
 become  ($'=d/d\xi$):
$$\eqalign{\zeta ''&=\ez,\cr
\Sigma ''&=g Q^2 e^{2\nu}\sinh(2g\Sigma) ,\cr
\nu ''&= Q^2 e^{2\nu}\cosh(2g\Sigma) ,\cr}
\eqno(3.3)$$
with the constraint
$$\zeta '^2-\nu '^2-\Sigma '^2-e^{2\zeta}+ Q^2 e^{2\nu}\cosh(2g\Sigma)=0.
\eqno (3.4)$$
Integrating the first equation in (3.3), the remaining  eqs. (3.3) and the
constraint (3.4)
are equivalent, respectively, to the equations
of motion and to the
Hamiltonian constraint derived from the Lagrangian:
$$L= {\nu'^2\over 2}+{\Sigma'^2\over 2}-V,$$
where the potential $V$ is given by
$$V=-{Q^2\over 2} e^{2\nu}\cosh(2g\Sigma).$$
This lagrangian describes two particles of mass equal to one moving on a
line and interacting through the potential $V$.
The system (3.3) can be solved exactly for $g^2=0,1,3$.
When $g^2=0$ the solutions
are the  RN solutions of
general relativity.
For $g^2=1,3$ the equivalent dynamical systems represent the Toda molecule
$SU(2)\times SU(2)$ and $SU(3)$ respectively [11]. We treat the two cases
separately.
\smallskip
\noindent
\leftline {\tscors {3.1 $\quad g^2=1.$}}

After some manipulations, we find a three-parameter class of solutions
describing asymptotically flat
black holes with a regular event horizon:
$$\eqalign{e^{2\Sigma}&=e^{2\Sigma_\infty}\left(1+{2\sigma \over r}\right),
\cr
ds^2&=-{(r-r_-)(r-r_+)\over{r(r+2\sigma)}}\,d
t^2+{r(r+2\sigma)\over (r-r_-)(r-r_+)}d
r^2+{r(r+2\sigma)}d\Omega^2.\cr}\eqno(3.5 )$$
The constants $\sigma$ and  $\Sigma_\infty$ are respectively the scalar charge
and   the
asymptotic
value of the field $\Sigma$. They are  defined through the asymptotic
behavior
$\Sigma \to \Sigma_\infty +\sigma/ r.$
The constants $r_+$ and $ r_-$ are related to the mass,
magnetic
and scalar charges of the black hole  through:
$$r_\pm= M-\sigma \pm\sqrt {M^2+\sigma^2-Q^2\cosh(2\Sigma_\infty)}\,.\eqno
(3.6)$$
The parameters $M$, $\sigma$, $Q$ and $\Sigma_\infty$ are not independent
but are constrained by
$$\sigma=-{Q^2\over2M}\sinh2\Sigma_\infty.\eqno(3.7)$$

The duality symmetry $\Sigma\to -\Sigma$ of the action (2.1) acts on the
space of the solutions transforming $\sigma\to -\sigma$ and
$\Sigma_\infty\to-\Sigma_\infty$. Therefore, we can restrict our
discussion to the case $\sigma>0$, $\Sigma_\infty<0$.
The solutions (3.5) describe  black holes only
for
 $${M^2+\sigma^2-Q^2\cosh2\Sigma_\infty}\geq 0.\eqno(3.8)$$
We have a curvature singularity at $r=0$ shielded by an inner (Cauchy)
horizon at $ r=r_-$ and by an  outer (event) horizon at
$r=r_+$.
The equality in (3.8) holds
in the extremal limit, $r_+= r_-$. Using (3.7)
the condition of  extremality  can be written in another form:
$$M^2-\sigma^2-Q^2=0.$$This means that in the extremal limit the
gravitational attraction is balanced by the repulsive forces of the
magnetic and scalar fields.
The solutions (3.5) represent a three-parameter class of solutions
generalizing the well-known RN solutions, to which they
reduce  when we have a self-dual configuration for the field $\Sigma$,
i.e for $\sigma=\Sigma_\infty=0$.
The presence of the scalar charge modifies  the area of the spheres $r=const,
t=const$ in the  RN solution, but the main features of the
latter are still preserved. In
particular, one can easily verify, performing the usual Kruskal extension
of the solutions (3.5), that the causal structure of the spacetime
(the Penrose diagram)
is exactly the same as in the  RN case.
The closed resemblance with the  RN solution is also
confirmed by the calculation of the thermodynamical parameters associated
with
 the  black hole.
For the temperature and the entropy of the hole we find:
$$T= {1\over 4\pi} {r_+-r_-\over r_+(r_+ +2\sigma)}, \quad
S= \pi  r_+(r_+ +2\sigma).$$
These formulae differ from the ones for the RN case only in
the area of the spheres with $r=const,
t=const$.

It is also interesting to compare our solutions to the GHS solutions.
The action (2.1) becomes the GHS action in the weak coupling regime
$\Sigma \to -\infty$ (due to the duality invariance of the action (2.1)
also in the strong coupling regime $\Sigma \to \infty$).
This regime can be studied by considering the behavior of the solutions
(3.5) for $\Sigma_\infty \to -\infty$.
Using the eqs. (3.6) and (3.7), one easily finds that in this limit
the inner horizon disappears. After the translation $r\to r-2\sigma$, one
has  $r_-=2 \sigma= Q^2 \exp(-2\Sigma_\infty)/2M$,
 $r_+=2M$ and the solutions (3.5) become
$$e^{-2\Sigma}=e^{-2\Sigma_\infty}\left(1-{ Q^2\over 2M} {e^{-2\Sigma_\infty}
\over r}\right),$$
$$ds^2=-\left(1- {2M\over r}\right )d
t^2+\left (1- {2M\over r}\right )^{-1}d
r^2+r\left (r-{ Q^2\over 2M} e^{-2\Sigma_\infty}\right)d\Omega^2,$$
which is   the GHS solution (the redefinition $Q^2\to 2Q^2$
is needed to match the conventions of ref. [1]).
It is important to notice  that the strong (or weak) coupling regime
 we consider here, is slightly different from
the strong (or weak) string coupling region that one usually considers
in the GHS model. Normally, one takes the
asymptotic value $ \Sigma_\infty$ of the dilaton constant, the
strong coupling region is then obtained by considering a spacetime region near
the singularity where the dilaton diverges. Our strong coupling regime
is obtained just by acting on the parameter $\Sigma_\infty$, without any
reference to a particular spacetime region. In particular, this means that
in the strong coupling regime the theory is strong coupled even in the
asymptotically flat, $r\to \infty$ region.
The lesson to be learned here is that the parameter $\Sigma_\infty$ is crucial
 to understand fully the parameter-space of the black hole
 solutions in string effective theory. The relevance of this parameter is
 also evident if one  considers it as the vacuum expectation value of the
 dilaton. It is well-known that at the tree level in the string
 perturbation theory this parameter is undetermined, though nonperturbative
 effect may fix it to some value, and that it is related to different
 possible string vacua.
\smallskip
\noindent
\leftline{\tscors{3.2 $\quad g^2=3.$}}

Also here one can use the equivalent dynamical system given by
the Toda molecule $SU(3),$ to find the solutions of the field equations
(3.3). This case has been treated by several authors (see for
example [12] and references therein), here we will
use a form of the solutions that is particularly suitable for our purposes.
The asymptotically flat black hole solutions can be written in the form:

$$e^{2\sqrt{3}\Sigma}={e^{2\sqrt{3}\Sigma_\infty}}\left({P_2(r)\over{P_1}(r)}
\right)^{3/2},\eqno(3.9)$$
$$ds^2 =-{(r-M)^2-q^2\over\sqrt{P_1(r)P_2(r)}}\,d
t^2 +{\sqrt{P_1(r)P_2(r)}\over(r-M)^2-q^2}\,d
r^2 +\sqrt{P_1(r)P_2(r)}\,d\Omega^2,\eqno(3.10)$$

where:
$$P_1(r)=\left(r-{\sigma\over \sqrt{3}}\right)^2- {Q^2\sigma
e^{2\sqrt{3}\Sigma_\infty}\over(\sigma - \sqrt{3}M)},$$
$$P_2(r)=\left(r+{\sigma\over \sqrt{3}}\right)^2- {Q^2\sigma
e^{-2\sqrt{3}\Sigma_\infty}\over(\sigma + \sqrt{3}M)},$$
$$q^2= M^2+\sigma^2- Q^2\cosh(2\sqrt{3}\Sigma_\infty).$$
The parameters $M$, $\sigma$ and $\Sigma_\infty$ appearing in the previous
equations are  respectively the mass, the scalar charge and the
asymptotic value of the field $\Sigma$. (3.9) and (3.10) are   solutions
of the
field equations  only if these parameters are related to the magnetic
charge by
$$Q^2\left( e^{2\sqrt{3}\Sigma_\infty}(\sigma + \sqrt{3}M)+
e^{-2\sqrt{3}\Sigma_\infty}(\sigma - \sqrt{3}M)\right)=
{4\over 3} \sigma(\sigma^2-3M^2).\eqno(3.11)$$
We have a three-parameter class of solutions. In the same way as
for $g^2=1$ the duality  symmetry of the action (2.1) relates solutions with
opposite signs of
$\sigma$ and $\Sigma_\infty$.  We will therefore consider only solutions with
$\sigma>0$, $\Sigma_\infty<0$.
The solutions (3.9), (3.10) represent black holes only
for
 $$\eqalign{M^2&+\sigma^2-Q^2\cosh (2\sqrt{3}\Sigma_\infty)\geq 0,\cr
 e^{2\sqrt{3}\Sigma_\infty}&\leq \bigl( {\sqrt{3}M-\sigma\over
 \sqrt{3}M+\sigma}\bigr)^{3/2}, \quad \sigma<\sqrt{3}M.\cr}\eqno(3.12)$$
For these values of the parameters, $P_1(r)$ is always positive whereas
$P_2(r)$ has two zeroes $r_1,r_2$, with $r_1<r_2$. The solutions are defined
for $r>r_2$. The spacetime has two horizons at $r_{\pm}= M\pm q$, the
inner of which screens the timelike singularity at $r=r_2$ to any
observer in the exterior region, just like  in a RN black
hole.
The Carter-Penrose diagram of the spacetime is therefore the same as in the
RN case. When the equalities in eqs. (3.12) hold, the
black hole becomes extremal and the spacetime has the causal structure
of the extremal RN black hole. The solutions (3.10) reduce to the RN
solutions in the  self-dual configuration of the field $\Sigma$, i.e. when
the scalar charge $\sigma$  vanishes, which  implies from (3.11) also
$\Sigma_\infty=0$. In the weak (or strong) coupling regime
$\Sigma_\infty\to -\infty$ the solutions reduce to that found by GHS in
ref. [1] for the model with coupling function $f= \exp(-2\sqrt{3}\Sigma)$.
In this regime the solutions (3.9),(3.10) and the constraint (3.11)
become respectively (we rescale $Q^2\to 2 Q^2$ to match the conventions
of ref. [1])
$$e^{-2\sqrt{3}\Sigma}=e^{-2\sqrt{3}\Sigma_\infty}
\left(1-{r_-
\over r}\right)^{3/2},$$
$$ ds^2=- \left(1-{r_+\over r}\right) \left(1-{r_-\over r}\right)^{-1/2}d
t^2 +\left(1-{r_+\over r}\right)^{-1} \left(1-{r_-\over r}\right)^{1/2}d
r^2 +r^2\left(1-{r_-\over r}\right)^{3/2}d \Omega^2,$$
$$ Q^2 e^{-2\sqrt{3}\Sigma_\infty}= {r_+r_-\over 4},$$
with the mass of the solutions given by $2M=r_+-r_-/2$.

To conclude this section let us calculate the thermodynamical parameters
associated with the black hole solution (3.10). We have for the temperature
and the entropy:
$$T={1\over 2\pi}{r_+-r_-\over \sqrt{P_1(M+q)P_2(M+q)}},$$
$$S= \pi \sqrt{P_1(M+q)P_2(M+q)}\,.$$

\beginsection 4. Dyonic black holes.
\smallskip

In the previous section we have seen that the implementation of a
S-duality symmetry at the string effective action level changes
drastically the structure of the black hole solutions with respect to
those of  the GHS model.
We found  not only that the black
hole solutions of the action (2.1) have the  causal structure of the
RN solutions but also that the GHS solutions emerge as
an approximation in the weak (or strong) coupling regime.
Up to now nobody has
shown that S-duality is a symmetry of string
theory, it remains just a conjecture. One could therefore object that our
results are just a peculiarity of our model (2.1) and that the true
description of charged black holes in string effective theory is that
given by the GHS model.
In the following we shall show that our model (2.1) can be used to
generate black hole solutions of the GHS model with both magnetic and
electric charges. Surprisingly enough, these dyonic solutions turn out
to be similar to our solution (3.5).

Let us consider the following action:
$$A=\int d^4x\sqrt{-g}\biggl(R-b(\nabla
\Sigma)^2-{f(\Sigma)}F^2\biggr).\eqno(4.1)$$
where $b$ is an arbitrary parameter.
One can  shown that the spherically symmetric solutions
$$d s^2=-e^{2\nu}d
t^2+e^{2\lambda}dr^2+e^{2\rho}d\Omega^2,\eqno(4.2)$$
of the action (4.1)
with a purely magnetic configuration for the EM field and  coupling
function $f$ given as follows:
$$\eqalign{f&=h +{1\over h},\cr
F&=Q_M \sin\theta{d\theta}\wedge{d\varphi},\cr}\eqno(4.3)$$
coincide with spherically symmetric solution of the action (4.1)
with the following dyonic configuration  for the EM field and coupling
function $f$:
$$\eqalign{f&=h,\cr
F&=-{Q_M\over h}e^{\nu+\lambda-2\rho} {dt}\wedge{d r}+Q_M{\sin\theta}
{d\theta}\wedge{d\varphi}.\cr}\eqno(4.4)$$
In fact, the field equations for the EM field:
$$\nabla_\nu\left( f F^{\nu\mu}\right)=0$$
are identically satisfied both for $f$, $F$ given by (4.3) or
by (4.4). The field  equations for the metric and the field $\Sigma$:
$$\eqalign{\rie &=b\nabla_\mu
\Sigma\nabla_\nu\Sigma+2f\left(F_{\mu\rho}F_\nu^\rho-
{1\over 4}F^2g_{\mu\nu}\right),\cr\nabla^2\Sigma &={1\over
2b}{df\over d \Sigma} F^2\cr}\eqno (4.5)$$
remain invariant inserting for $f$ and $F$ the expressions (4.3) or (4.4).
Using this equivalence, we can generate dyonic solutions for the GHS theory,
 i.e
for $f=h=\exp(-2\Sigma)$, from the magnetic solution of the model (2.1)
with $g^2=1$. These dyonic solutions are given by (3.5), with $Q^2=
2Q^2_M$, and by the EM form:
$$F=-{Q_M\over r^2}e^{2\Sigma_\infty} {dt}\wedge{d r}+Q_M{\sin\theta}
{d\theta}\wedge{d\varphi}.\eqno(4.6)$$
The electric charge $Q_E$ of the solutions is related to the magnetic
charge by
$$Q_E= Q_M e^{2\Sigma_\infty}.\eqno(4.7)$$
It is also evident from the construction of the solutions that the
S-duality symmetry of the action (2.1) is
related to the EM duality of the field equations of GHS  model.
In fact,  the field equations  of the latter are invariant under the
EM duality transformation:
$$\Sigma\to -\Sigma , \quad F\to hF^*.$$
These dyonic solutions of the GHS model have the  spacetime
structure of the solutions with $g=1$ discussed in sect. 3.
In particular, as pointed out is sect. 3, they are very similar to
the RN solutions and differ drastically from the purely
magnetic (or electric) solutions found in ref. [1].
Using eq. (4.7) and from the discussion of sect. 3, one finds that
the purely magnetic (electric) solutions of the GHS model can
be found as a limit of the dyonic ones in the weak (strong) coupling regime
$\Sigma_\infty\to -\infty$ ($\Sigma_\infty\to \infty$).
The dyonic solutions,
(4.6), (3.5) have been already found by
Gibbons [12] and Gibbons
and Maeda [4]. In the latter paper it was also pointed out that
the solutions have the
same Penrose diagram as the RN solutions. Furthermore, the solutions in an
explicit form and with a nonvanishing  asymptotic value of the scalar field,
 as (3.5), have been
lately found in [13]. As far as the case with $g^2=3$ is
concerned, one can show, using arguments similar to those used for
$g^2=1$,
 that the solutions (3.9), (3.10) also represent  Kaluza-Klein black holes
with $Q_E= Q_M \exp(2\sqrt{3}\Sigma_\infty).$ These black hole
solutions are similar to the Kaluza-Klein black hole solutions found  in
refs. [12,14].

\beginsection 5. Conclusions.
\smallskip

In this paper we have found charged black hole solutions of a string
effective theory invariant under S-duality transformations. The picture
that has emerged  from the study of these solutions is rather intriguing:
the black hole solutions are similar to the RN solutions of general
relativity,
in particular they share with the last-named  the causal structure, reduce to
the RN case for  self-dual configurations of the dilaton
 and to the GHS
black holes in the weak (strong ) coupling regime $\Sigma_\infty\to - \infty$
($\Sigma_\infty\to \infty$ ).
We have seen that this picture of charged black holes emerges also
in the context of the GHS model if one considers dyonic configurations for
the EM field.
In view of these results one is led to
conclude that the description of charged black holes in string theory can
be reconciled with the RN description if one goes beyond the
tree-level approximation for the coupling function $f$ or, even in this
approximation, if one considers black holes with both electric and
magnetic charges. The statement,
affirming that the RN solutions are not an approximate solutions of string
theory is, therefore, only true at the tree level in the string
perturbation theory, where the coupling function $f=\exp(-2\Sigma)$ and
if one considers only purely magnetic (or purely electric) EM field
configurations.
As we have shown considering a S-duality model, this statement can be
invalidated both with a different choice for $f$ or with dyonic
configurations for the EM field.

Though reasonable our description of charged black holes
is still incomplete and far from giving a definitive  answer to the
question about the true description of charged black holes in string
theory.
Our
results rely heavily on the existence of a   S-duality symmetry
of the string effective action, which exchanges strong and weak string
 couplings.
This is just a conjectured symmetry
of string
theory and one cannot be sure that it really holds.
On the other hand  the dyonic solutions we have found for the GHS
model
 represent a special case of the generic dyonic solution of this model.
We are therefore not allowed to draw general conclusions from this
particular case, without having full control of the general solution.

Even though one
could prove that S-duality is a symmetry of string theory, it is not
evident a priori that our choice
for the coupling function $f$ is even a good approximation to the exact
S-duality invariant coupling function. Apart from the fact that the
coupling function $f$ has the form of a series of powers in the string
coupling function $g_s^2=\exp(2\Sigma)$ and that the genus-$n$
string-loop contributions contain the factor $g_s^{2(n-1)}$, little is
known about the exact form of $f$. However, the main features of the
solutions we found (existence of RN solutions for self-dual
configurations of the dilaton, existence of a strong or weak coupling
regime in which the solutions have the GHS form) seem to be consequence
of the symmetry of the model, namely the S-duality symmetry.
One would therefore expect that these main features do not depend on the
particular functional form of $f$.
\smallskip

\beginack

We thank S. Mignemi  for useful comments.
\smallskip

\beginref

\ref [1]  D. Garfinkle, G.T. Horowitz and A. Strominger, \PRD {\bf 43} (1991)
 3140.

\ref [2] W. Israel, \CMP {\bf 8}, 245 (1968); J. Chase, \CMP {\bf 19} (1970)
276.

\ref [3] E. S. Fradkin, A. A. Tseytlyn, \PLB {\bf 158}, 316 (1985);
C. G. Callan, D. Friedan, E. J. Martinec and M. S. Perry, \NPB {\bf 262}
(1985)
593.

\ref [4]   G.W. Gibbons and K. Maeda, \NPB {\bf 298} (1988)  748.

\ref [5] G. T. Horowitz, {\tscors The Brill Festschrift}, B. L. Hu and T.
   A. Jacobson, eds, Cambridge University Press (1993);
   C.F.E. Holzhey and F. Wilczek, \NPB  {\bf 380} (1992)  447;
  A. Shapere, S. Trivedi and F. Wilczek, \MPLA  {\bf 29} (1991) 2677;
   A. G. Agnese, M. La Camera \PRD {\bf 49} (1994) 2126.

\ref [6] M. Cadoni, S. Mignemi \PRD {\bf 48} (1993) 5536.

\ref [7] M . Cveti\v c, A. A. Tseytlin, \NPB {\bf 416} (1994) 137.

\ref [8] A. Font, L. E. Ib\'a\~ nez, D. L\"ust and F. Quevedo, \PLB {\bf
249} (1990) 35; J.H. Schwarz, A. Sen, \PLB {\bf 312} (1993) 105; \NPB
{\bf 411}  (1994) 35.

\ref [9] L. J. Dixon, V. S.  Kaplunovsky, J. Louis, \NPB {\bf 355}
 (1991) 649.

\ref [10] M. K. Gaillard, B. Zumino, \NPB {\bf 193}  (1981) 221.

\ref [11] M.A. Olshanetsky and A.M. Perelomov, \PRP {\bf 71} (1981) 313.

\ref [12] G.W. Gibbons, \NPB {\bf 207} (1982) 337.

\ref [13] R. Kallosh, A. Linde, T. Ort\'in, A. Peet, \PRD {\bf 46} (1992) 5278.

\ref [14] G.W. Gibbons and D.L. Whiltshire, \ANP {\bf
167}  (1986) 201.

\endref
\end